\documentclass[a4paper]{panl}
\usepackage{cite}
\usepackage{wrapfig}
\usepackage{graphicx}
\usepackage{amssymb}
\usepackage{amsfonts}
\usepackage{amsmath}
\usepackage{longtable}
\usepackage{rotating}
\usepackage{lscape}
\usepackage{epsfig}
\usepackage{multirow}

\originalTeX

\newcommand{\AGeV}{$A$GeV}
\newcommand{\fig}[1]{Fig.~\ref{#1}}
\newcommand{\eq}[1]{Eq.~(\ref{#1})}
\newcommand{\tb}[1]{Table~\ref{#1}}

\begin{document}

\title{Estimating non-flow effects in measurements of directed flow of protons with the HADES experiment at GSI}
\maketitle
\authors{M.~Mamaev$^{ a,}$\footnote{E-mail: mam.mih.val@gmail.com}, O.~Golosov$^{ a,}$\footnote{E-mail: oleg.golosov@gmail.com} and I.~Selyuzhenkov$^{ a,b,}$\footnote{E-mail: ilya.selyuzhenkov@gmail.com}}

\vspace{0.1cm}
\authors{for the HADES Collaboration}
\from{$^{a}$~National Research Nuclear University MEPhI, 115409 Moscow, Russia}
 \vspace{-0.4cm}
\from{$^{b}$~GSI Helmholtzzentrum für Schwerionenforschung GmbH, 64291 Darmstadt, Germany}

\begin{abstract}
\vspace{0.2cm}
Centrality dependence of the directed flow of protons in Au+Au collisions at the beam energy of 1.23\AGeV{} collected by the HADES experiment at GSI is presented.
Measurements are performed with respect to the spectators plane estimated using the Forward Wall hodoscope.
Biases due to non-flow correlations and correlated detector effects are evaluated.
The corresponding systematic uncertainties are quantified using estimates of the spectators plane from various forward rapidity regions constructed from groups of Forward Wall channels and protons reconstructed with the HADES tracking system.

\end{abstract}

\section*{Introduction}

Momentum anisotropy of the particle production in a heavy-ion collision is a response of the strongly interacting matter to the spatial asymmetry of the initial geometry. 
A comparison of the measured azimuthal anisotropies (collective flow) with theoretical calculations allows to extract information about the created matter, such as its equation of state.
The multi-differential measurement of the anisotropic flow of protons and light nuclei has been recently reported by the HADES Collaboration~\cite{Adamczewski-Musch:2020iio}.

The collective flow is quantified with the coefficients $v_{n}$ in a Fourier decomposition of the azimuthal distribution of produced particles relative to the reaction plane spanned by the impact parameter and the beam direction~\cite{Voloshin:2008dg}:
\begin{equation}
    \rho( \phi - \Psi_{RP} ) = \frac{1}{2\pi} [ 1+2\sum_{n=1}^\infty v_n \cos(\phi-\Psi_{RP}) ],
    \label{eq:particle_distribution}
\end{equation}
where $\Psi_{RP}$ is a reaction plane angle.
The $v_n$ coefficients can be calculated as 
\begin{equation}
v_n=\langle \cos n (\phi - \Psi_{RP}) \rangle,
\end{equation}
where the angle brackets denote an average over all particles in an event and all events.
The first coefficient of the  Fourier expansion, $v_1$, is called directed flow.

The reaction plane can be estimated from the transverse deflection of the collision spectators (nucleons and fragments).
The collision spectators can not be identified uniquely and a contamination from produced nucleons and fragments at forward rapidity may introduce biases in the spectator plane determination.

Experimental measurements of the directed flow are performed using correlations between the azimuthal angle of produced particles (protons) and the spectators symmetry plane angle estimated from azimuthal asymmetry of particles emitted at forward rapidity.
The contamination at forward rapidity from non-spectator particles may introduce additional (non-flow) correlations and biases the extracted $v_1$ values.
Among the sources of non-flow correlations are (for discussion see e.g. \cite{Borghini:2000cm, Voloshin:2008dg}):
(a) conservation of total (transverse) momentum in the collision;
(b) resonance decays
and (c) short-range correlations (femtoscopy, cluster decays).
Detector effects can also introduce correlated biases to the flow measurements.
Some examples are: reconstructed split (merged) tracks from a single (pair of) particle(s), or double (multiple) hits from the same particle in modules of the segmented detectors. 

The influence of non-flow and correlated detector biases can be reduced using multi-particle correlations~\cite{Bilandzic:2010jr, Bilandzic:2013kga} or by introducing a rapidity separation between correlated particles~\cite{Voloshin:2006wi}.
In these proceedings the evaluation of systematic uncertainties in the measurement of the proton directed flow due to non-flow and correlated detector effects is presented.

\section*{The HADES experiment and analyzed data sample}

\begin{figure}[h]
    \centering
    \includegraphics[width=0.6\textwidth]{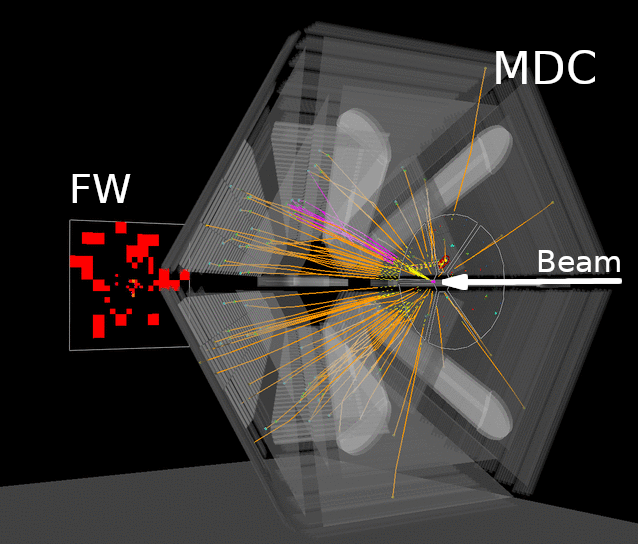}
    \caption{Event display of the HADES experiment. Reconstructed tracks in the Multiwire Drift Chambers (MDC) are displayed as colored lines.
    Modules of the Forward Wall hodoscope (FW) as a non-zero signal are shown with red boxes.
    }
    \label{fig:event_display}
\end{figure}

The High Acceptance Di-Electron Spectrometer (HADES) is a fixed target experiment at SIS18 accelerator in GSI, Darmstadt~\cite{Agakishiev:2009am}.
The relevant sub-systems for the proton flow measurements are: MDC (Multiwire Drift Chambers), TOF (Time Of Flight) detector, RPC (Resistive Plate Chambers) and FW (Forward Wall) hodoscope.
Tracks associated with produced charged particles are reconstructed with the MDC.
The  TOF and RPC are used for identification of protons.
Charges of the spectator fragments are measured with the FW.
Transverse distribution of the FW signals from charged spectator fragments is used to estimate the orientation of the spectators plane (for illustration, see HADES event display in~\fig{fig:event_display}).

Approximately one billion of Au+Au collisions at beam energy of 1.23\AGeV{} were recorded by HADES in the year 2012.
A subset of 10M Au+Au collisions for 0--40\% centrality is used in this analysis. 
Estimation of the collision centrality is based on the total amount of hits in the TOF and RPC detectors according to the procedure described in~\cite{Adamczewski-Musch:2017sdk}.
Events with the reconstructed vertex position $(x_v,y_v,z_v)$ outside the target volume $\sqrt{x_v^2+y_v^2}>3$~mm and $z_v \notin (-60, 0)$~mm were rejected.
Events with a single vertex inside the target region were selected to suppress a contamination from pileup.

Only tracks with fit quality $\chi^2<100$ were used in the analysis.
The selection of primary particles is based on the distance of the closest approach ($DCA$) of the associated track to the collision vertex.
Only tracks with $DCA$ within $(-10,10)$~mm were used in the analysis.

\section*{Flow observables}

The observables for $v_1$ coefficient can be written in terms of flow vectors.
For each particle in the event a vector in the transverse plane $u_1$ is calculated:
\begin{equation}
    u_1 = e^{i\phi},
\end{equation}
where $\phi$ is the azimuthal angle of the particle's momentum.
A $Q_1$-vector is defined as a sum of $u$-vectors over a group of particles:
\begin{equation}
    Q_1 = (1/M)\sum_{k=1}^{M} u_{1,k} = X+iY = |Q_1| e^{i\Psi_{EP,1}},
    \label{eq:q_vector}
\end{equation}
where $M$ is the number of particles in the group, $X$ and $Y$ are the components of the $Q_1$-vector.
The azimuthal angle $\Psi_{EP,1}$ is called an event plane angle and it gives an estimate of the reaction plane angle.

The observable for the directed flow can be measured using correlations between $u_1$ and $Q_1$ vectors:
\begin{equation}
    v_1 = \frac{\langle u_1 Q_1 \rangle}{R_1},
\end{equation}
where $R_1$ is the resolution correction factor.
In the present analysis the resolution correction factor is calculated using the method of three sub-events given by the equation:
\begin{equation}
    R_1\{a(b,c)\} = \sqrt{\frac{ \langle Q_1^{a} Q_{1}^{b} \rangle \langle Q_1^a Q_1^c \rangle }
    {\langle Q_1^b Q_1^c \rangle}},
    \label{eq:res_3se}
\end{equation}
where indexes "a", "b" and "c" denote groups of particles (sub-events) in which $Q_1$-vector is calculated separately.

\begin{figure}
    \centering
    \includegraphics[width=0.7\textwidth]{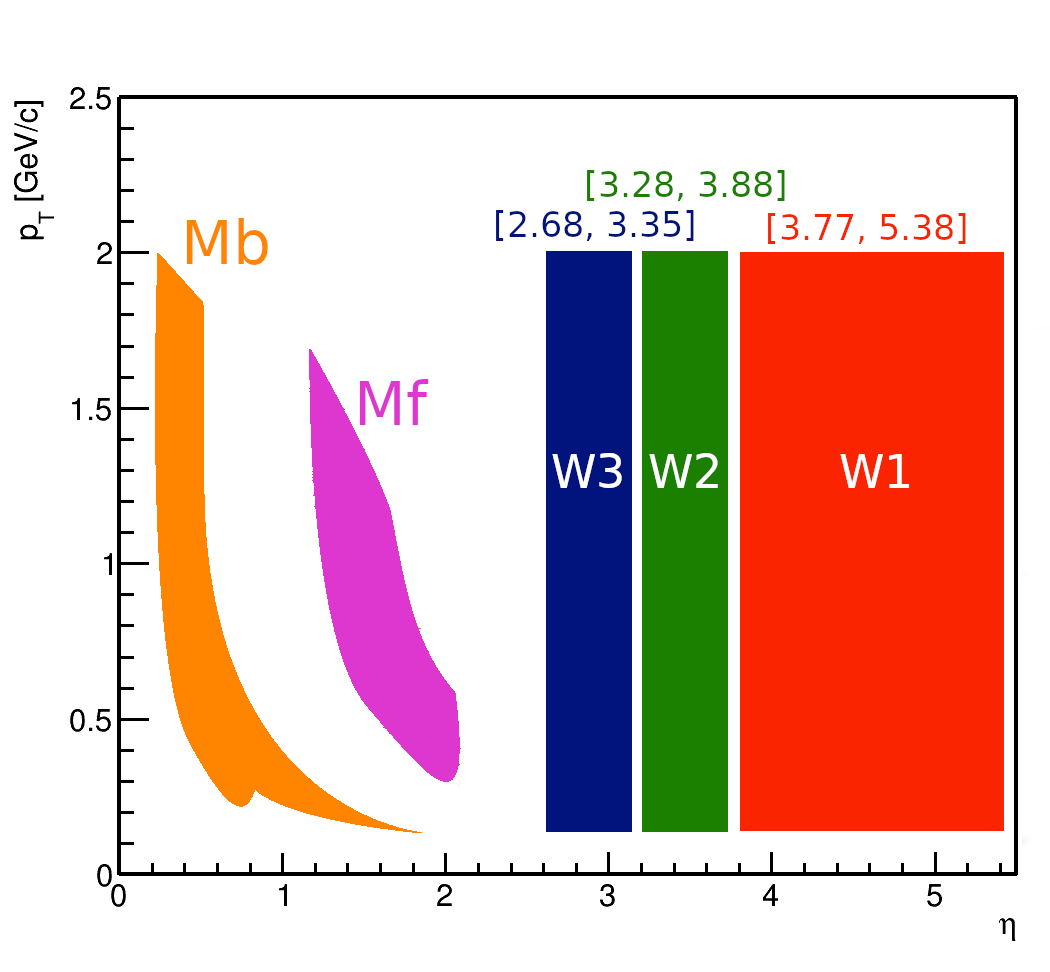}
    \caption{Schematic view of the pseudorapidity (for MDC tracks also $p_T$) coverage of the MDC and FW sub-events used for the estimation of the biases due to non-flow correlations and correlated detector effects.}
    \label{fig:flow_se}
\end{figure}

Different sub-events (in total five) were used to obtain estimates of the FW resolution correction factor (see~\fig{fig:flow_se}).
Two sub-events were used from proton tracks reconstructed and identified with the MDC, TOF, and RPC: (Mf) $0.35<y_{cm}<0.55$ and (Mb) $-0.55<y_{cm}<-0.35$, where $y_{cm}$ is the center-of-mass rapidity.
Modules of the FW were divided into three groups according to the ranges of pseudorapidity in the laboratory frame $\eta$:
(W1) $3.77 < \eta < 5.38$; (W2) $3.28 < \eta < 3.88$; and (W3) $2.68 < \eta < 3.35$.
One should note that due to the rectangular shape of the FW modules there is a small overlap in the pseudorapidity coverage between the FW sub-events.

The $Q_1$ vectors for the MDC subevents are calculated according to~\eq{eq:q_vector}.
The $Q_1$ vectors for the FW are obtained as follows:
\begin{equation}
Q_1 = \sum_{k=1}^{N} E_k e^{i\varphi_k} / \sum_{k=1}^{N} E_k,
\end{equation}
where $\varphi$ is the azimuthal angle of the $k$-th FW module, $E_i$ is its signal, and $N$ is the total number of modules with non-zero signal.

\section*{Biases due to non-flow correlations and correlated detector effects}

The azimuthal detector non-uniformity may bias the orientation and magnitude of the $Q$-vectors and result in a difference between correlations with the $X$~and~$Y$ $Q$-vector components. 
These effects can be corrected with the data-driven correction procedure described in~\cite{Selyuzhenkov:2007zi}.
The procedure implies recentering, twist and rescaling steps of the $Q$-vector corrections.
In the present analysis the non-uniformity corrections differential in $p_T$, $y_{cm}$, and centrality were applied using the QnTools software framework~\cite{QnTools:2020}, which incorporates the QnCorrections framework originally developed for the ALICE experiment at the LHC~\cite{QnGithub:2015}. 
The configuration of the framework used in the present analysis is shown in~\tb{table:qn_tools_config}.

\begin{table}[h]
\begin{tabular}{|c|l|l|l|l|c|}
\hline
\multicolumn{1}{|l|}{\textbf{flow vector}}
& \textbf{weight} 
& \textbf{norm}
& \textbf{\begin{tabular}[c]{@{}c@{}}correction\\ axes\end{tabular}}
& \textbf{\begin{tabular}[c]{@{}c@{}}correction \\steps\end{tabular}}
& \textbf{stat. errors} \\ \hline
\begin{tabular}[c]{@{}c@{}}$u_1$ [-0.25,-0.15]\\ Mf~[~0.35,~0.55]\\ Mb~[-0.55,-0.35]\end{tabular} 
& \multicolumn{1}{c|}{1}
& \multirow{3}{*}{\begin{tabular}[c]{@{}l@{}}sum of \\ weights\end{tabular}} 
& \begin{tabular}[c]{@{}l@{}}$p_T$ [0.0, 2.0]\\ centrality [0,40]\end{tabular} 
& \begin{tabular}[c]{@{}l@{}}recentering\\ twist\\ rescaling\end{tabular} 
& \multirow{2}{*}{\begin{tabular}[c]{@{}c@{}}bootstrap\\ 100 samples\end{tabular}} \\ \cline{1-2} \cline{4-5}
FW sub-events
& \begin{tabular}[c]{@{}l@{}}module \\ charge\end{tabular}
&
& centrality [0,40]
& recentering
&\\ \hline
\end{tabular}
\caption{Configuration of the QnTools software package~\cite{QnTools:2020} used for the proton directed flow analysis and corrections for detector azimuthal non-uniformity.}
\label{table:qn_tools_config}
\end{table}

\begin{figure}[h]
    \centering
    \includegraphics[width=1\textwidth]{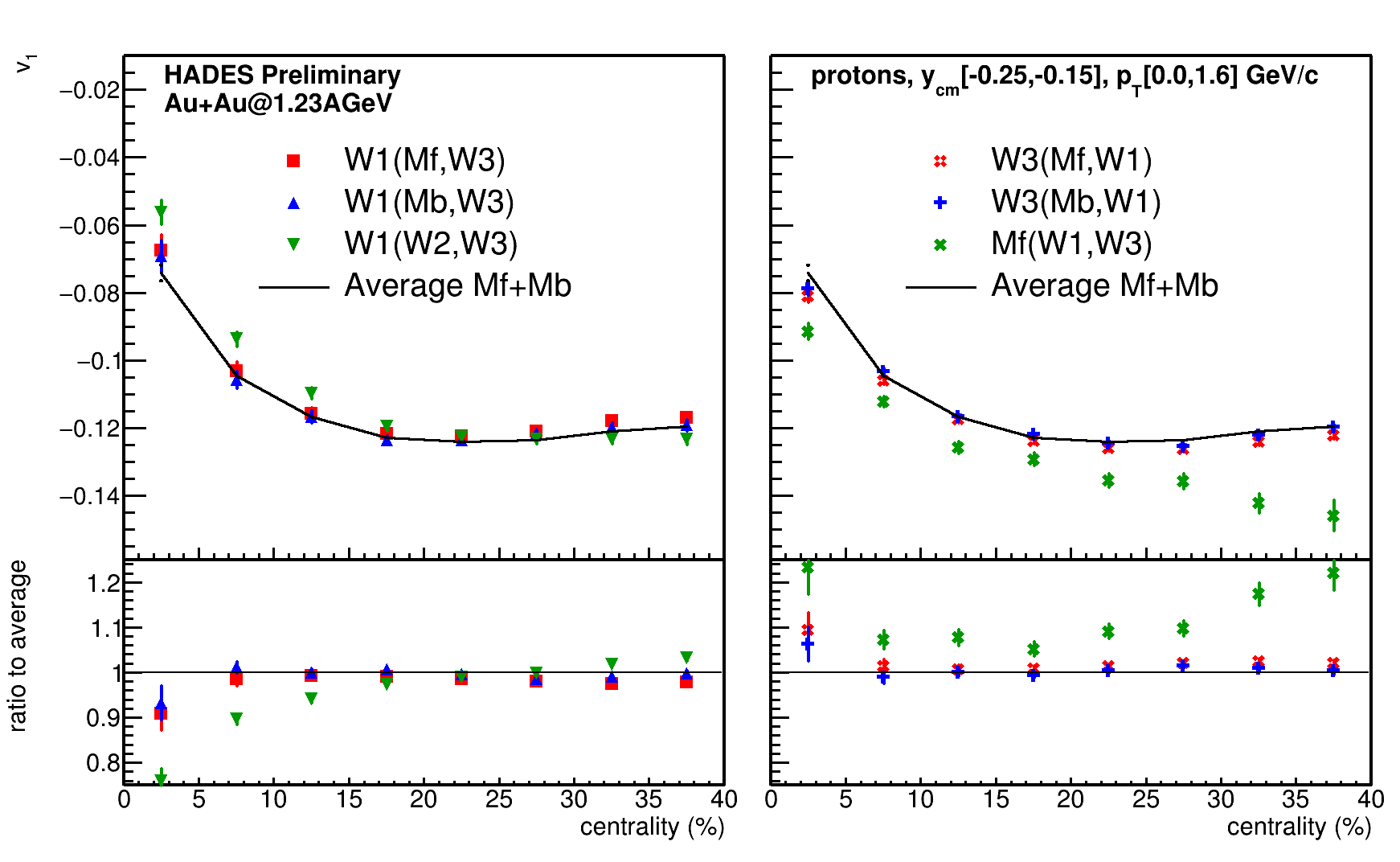}
    \caption{
    Directed flow, $v_1$, of protons as a function of collision centrality estimated with respect to (left) the inner and outer FW subevents, (right) Mf subevent.
    Different combinations of sub-events used for the $R_1$ calculations are represented with different markers.
    An average of the results for all combinations with rapidity-separated sub-events is shown as a black line.
    }
    \label{fig:results}
\end{figure}

The three sub-event method for the calculation of the resolution correction factor $R_1$ can be implemented using different combinations of five available $Q_1$-vectors (see~\eq{eq:res_3se}).
This allows to study the effect of non-flow correlations by using combinations of sub-events with different separation in rapidity.
In~\fig{fig:results} the results for directed flow of protons obtained using different combinations of $Q_1$-vectors are shown.
Results for $v_1$ obtained with W1(Mf,W3) and W1(Mb,W3) combinations as well as with W3(Mf,W1) and W3(Mb,W1) combinations
are consistent with each other within 2\% except of central collisions, where the difference increases up to 5\%.
This indicates that a rapidity separation of 0.5 between the FW sub-events is sufficient to suppress non-flow and correlated detector effects.

A difference of the proton $v_1$ values without (or with a small) rapidity separation, such as W1(W2,W3)  combination, is much larger.
Its is about 10\% in mid-central collisions and increases up to 20-30\% in most central collisions (see left panel of \fig{fig:results}). 
The directed flow calculated using $R_1$ from W1(W2,W3) has a smaller absolute value than with  W1(Mf,W3) or W1(Mb,W3) due to additional correlations between pairs of neighboring FW sub-events (i.e. W1 and W2 or W2 and W3).
This can happen  when spectator fragments passing through the FW material results simultaneously in a signal of two neighboring FW modules and yield to positive non-physical correlation between FW sub-events.
As a result the resolution correction factor is overestimated, while the flow value is underestimated.

The results for directed flow with respect to symmetry plane evaluated with protons from MDC are shown on the right panel of \fig{fig:results}. 
A difference between results obtained relative to the MDC and FW sub-events is more than 25\% in most central (0-5\%) and peripheral (>30\%) collisions.
This indicates that the contamination of spectator fragments by produced particles is decreasing with increasing rapidity.
A much larger variation of the proton $v_1$ results relative to the produced protons (estimated with the Mf sub-event) also suggests a significant effect due to the global momentum conservation~\cite{Borghini:2000cm, Voloshin:2008dg}.

\section*{Summary}
Directed flow $v_1$ of protons is measured in Au+Au collisions at the beam energy of 1.23\AGeV{} using the three sub-events method.
Systematic uncertainties due to non-flow correlations and correlated detector biases are evaluated using estimates of the spectators plane from various forward rapidity regions constructed from groups of the Forward Wall channels and protons reconstructed with the HADES tracking system.
It is found that a rapidity separation of 0.5 between the FW (spectator) sub-events is sufficient to suppress non-flow and correlated detector effects.
The overall deviation between results for $v_1$ obtained with rapidity separation and without it is about 20-30\% in most central and approximately 10\% in mid-central collisions.
A difference between results obtained relative to the MDC (produced protons) and FW (spectators) sub-events is more than 25\% in most central (0-5\%) and peripheral (>30\%) collisions.
This observation reflects a significant non-flow contribution (resulting from the global momentum conservation) to the proton $v_1$ measurement relative to the MDC sub-events.

\section*{Acknowledgments}
The work is supported by
the Ministry of Science and Higher Education of the Russian Federation, Project ``Fundamental properties of elementary particles and cosmology'' No 0723-2020-0041,
the Russian Foundation for Basic Research (RFBR) funding within the research project no. 18-02-40086,
the European Union‘s Horizon 2020 research and innovation program under grant agreement No. 871072,
the National Research Nuclear University MEPhI in the framework of the Russian Academic Excellence Project (contract no. 02.a03.21.0005, 27.08.2013).

\bibliographystyle{pepan}
\bibliography{references}

\end{document}